\documentclass[superscriptaddress,aps,pra,twocolumn,nofootinbib,babel]{revtex4-1}
\usepackage{graphicx}
\usepackage{amsmath}
\usepackage{amssymb}
\usepackage{comment}
\usepackage{placeins}
\usepackage{rotating}
\usepackage[caption=false]{subfig}
\usepackage[colorlinks]{hyperref}
\usepackage{hypcap}
\usepackage[english]{babel}
\usepackage{color}
\usepackage[normalem]{ulem}

\newcommand{\eq}[1]{Eq.~\hyperref[eq:#1]{(\ref*{eq:#1})}}
\renewcommand{\sec}[1]{\hyperref[sec:#1]{Section~\ref*{sec:#1}}}
\newcommand{\app}[1]{\hyperref[app:#1]{Appendix~\ref*{app:#1}}}
\newcommand{\tab}[1]{\hyperref[tab:#1]{Table~\ref*{tab:#1}}}
\newcommand{\fig}[1]{\hyperref[fig:#1]{Figure~\ref*{fig:#1}}}
\newcommand{\figa}[2]{\hyperref[fig:#1]{Figure~\ref*{fig:#1}#2}}
\newcommand{\figx}[2]{\hyperref[fig:#1]{Figure~\ref*{fig:#1}(#2)}}
\newcommand{\thm}[1]{\hyperref[thm:#1]{Theorem~\ref*{thm:#1}}}
\newcommand{\lem}[1]{\hyperref[lem:#1]{Lemma~\ref*{lem:#1}}}
\newcommand{\cor}[1]{\hyperref[cor:#1]{Corollary~\ref*{cor:#1}}}
\newcommand{\defn}[1]{\hyperref[def:#1]{Definition~\ref*{def:#1}}}
\newcommand{\alg}[1]{\hyperref[alg:#1]{Algorithm~\ref*{alg:#1}}}
\newcommand{\specialcell}[2][c]{%
  \begin{tabular}[#1]{@{}c@{}}#2\end{tabular}}

\def\bra#1{\mathinner{\langle{#1}|}}
\def\ket#1{\mathinner{|{#1}\rangle}}

\newcommand\jp{{\vec{p}}}

\newcommand{\ignore}[1]{}

\newcommand{\be}{\begin{equation}}
\newcommand{\ee}{\end{equation}}
\newcommand{\ba}{\begin{eqnarray}}
\newcommand{\ea}{\end{eqnarray}}
\usepackage{times}

\newcommand{\jrf}[1]{{\color{black}{#1}}}
\newcommand{\jonny}[1]{{\color{black}{#1}}}

\DeclareMathOperator{\Tr}{Tr}

\input{Qcircuit}

\begin{document}

\title{Quantum autoencoders for efficient compression of quantum data}

\author{Jonathan Romero}
\affiliation{Department of Chemistry and Chemical Biology, Harvard University, Cambridge, Massachusetts 02138, United States}

\author{Jonathan P. Olson}
\affiliation{Department of Chemistry and Chemical Biology, Harvard University, Cambridge, Massachusetts 02138, United States}

\author{Alan Aspuru-Guzik}
\email[Corresponding author: ]{aspuru@chemistry.harvard.edu}
\affiliation{Department of Chemistry and Chemical Biology, Harvard University, Cambridge, Massachusetts 02138, United States}

\date{\today}

\begin{abstract}
Classical autoencoders are neural networks that can learn efficient low dimensional representations of data in higher dimensional space. The task of an autoencoder is, given an input $x$, is to map $x$ to a lower dimensional point $y$ such that $x$ can likely be recovered from $y$.  The structure of the underlying autoencoder network can be chosen to represent the data on a smaller dimension, effectively compressing the input. Inspired by this idea, we introduce the model of a quantum autoencoder to perform similar tasks on quantum data. The quantum autoencoder is trained to compress a particular dataset of quantum states, where a classical compression algorithm cannot be employed. The parameters of the quantum autoencoder are trained using classical optimization algorithms. We show an example of a simple programmable circuit that can be trained as an efficient autoencoder. We apply our model in the context of quantum simulation to compress ground states of the Hubbard model and molecular Hamiltonians.
\end{abstract}

\maketitle

\section{Introduction}\label{sec:intro}

Quantum technologies, ranging from quantum computing to quantum cryptography, have been found to have increasingly powerful applications for a modern society.  Quantum simulators for chemistry, for example, have been recently shown to be capable of efficiently calculating molecular energies for small systems \cite{omalley.PRX.6.031007.2016}; the capability for larger scale simulations promises to have deep implications for materials design, pharmacological research, and an array of other potentially life-changing functions \cite{Aspuru-Guzik.S.309.1704.2005}.  A limiting factor for nearly all of these applications, however, is the amount of quantum resources that can be realized in an experiment.  Therefore, for experiments now and in the near future, any tool which can reduce the experimental overhead in terms of these resources is especially valuable.

For classical data processing, machine learning via an autoencoder is one such tool for dimensional reduction \cite{ae1,ae2,BiamonteLloyd2016}, as well as having application in generative data models \cite{ae3}.  A classical autoencoder is a function whose parameters are optimized across a training set of data which, given an $(n+k)$-bit input string $x$, attempts to reproduce $x$.  Part of the specification of the circuit, however, is to erase some $k$ bits during the process.  If an autoencoder is successfully trained to reproduce $x$ at the output \jonny{at least approximately}, then the remaining $n$ bits immediately after the erasure (referred to as the \textit{latent space}) represent a compressed encoding of the string $x$.  Thus, the circuit ``learns" to encode information that is close to the training set.

In this paper, we introduce the concept of a quantum autoencoder which is inspired by this design for an input of $n+k$ qubits.  \jonny{Because quantum mechanics is able to generate patterns with properties (e.g. superposition and entanglement) that is beyond classical physics, a quantum computer should also be able to \textit{recognize} patterns that are beyond the capabilities of classical machine learning.}  Thus, the motivation for a \textit{quantum} autoencoder is simple; such a model allows us to perform analogous machine learning tasks for quantum systems without exponentially costly classical memory, for instance, in dimension reduction of quantum data.  A related work proposing a quantum autoencoder model establishes a formal connection between  classical and quantum feedforward neural networks \jonny{where a particular setting of parameters in the quantum network reduces to a classical neural network exactly} \cite{FFNN2016}. In this work, we provide a \jonny{simpler} model which we believe more easily captures the essence of an autoencoder, \jonny{and apply it to ground states of the Hubbard model and molecular Hamiltonians.} 

\begin{figure}
\includegraphics[width=8cm]{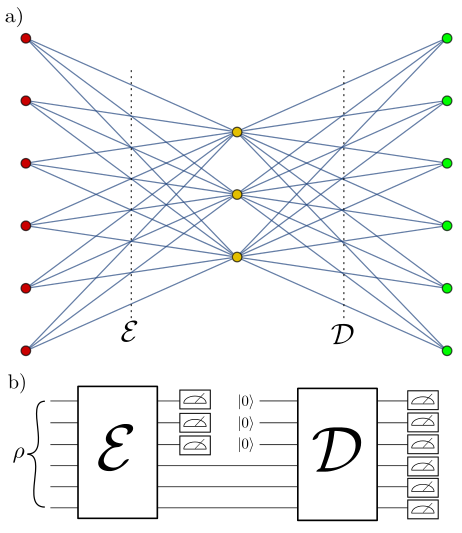}
\caption{a) A graphical representation of a 6-bit autoencoder with a 3-bit latent space.  The map $\mathcal{E}$ encodes a 6-bit input (red dots) into a 3-bit intermediate state (yellow dots), after which the decoder $\mathcal{D}$ attempts to reconstruct the input bits at the output (green dots). b) Circuit implementation of a 6-3-6 quantum autoencoder.} \label{fig:636}
\end{figure}

\section{Quantum Autoencoder Model}\label{sec:model}

In analogy with the model of classical autoencoders, the quantum network has a graphical representation consisting of an interconnected group of nodes. In the graph of the quantum network, each node represents a qubit, with the first layer of the network representing the input register and the last layer representing the output register. In our representation, the edges connecting adjacent layers represent a unitary transformation from one layer to the next.  Autoencoders, in particular, shrink the space between the first and second layer, as depicted in \fig{636}a.

For a quantum circuit to embody an autoencoder network, the information contained in some of the input nodes must be discarded after the initial ``encoding" $\mathcal{E}$.  We imagine this takes place by tracing over the qubits representing these nodes (in \fig{636}b, this is represented by a measurement on those qubits).  Fresh qubits (initialized to some reference state) are then prepared and used to implement the final ``decoding" evolution $\mathcal{D}$, which is then compared to the initial state.


The learning task for a quantum autoencoder is to find unitaries which preserve the quantum information of the input through the smaller intermediate latent space.  To this end, it is important to quantify the deviation from the initial input state, $\ket{\psi_i}$, to the output, $\rho_i^{out}$.  Here, we will use the expected fidelity \cite{Wilde2012} $F(\ket{\psi_i},\rho_i^{out})=\bra{\psi_i}\rho_i^{out}\ket{\psi_i}$.  We thus describe a successful autoencoding as one in which $F(\ket{\psi_i},\rho_i^{out})\approx 1$ for all the input states.

\begin{figure}[h]
\includegraphics[width=8.4cm]{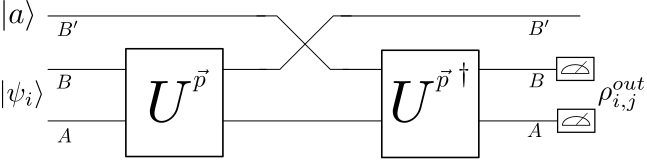}
\caption{A quantum autoencoder circuit.  The goal is to find $\jp$ such that the averaged $F(\ket{\psi}_i,\rho_{i,\jp}^{out})$ is maximized.} \label{fig:generalcircuit}
\end{figure}

A more formal description of a quantum autoencoder follows: Let $\{p_i,\ket{\psi_i}_{AB}\}$ be an ensemble of pure states on $n+k$ qubits, where subsystems $A$ and $B$ are comprised of $n$ and $k$ qubits, respectively.  Let $\{U^\jp\}$ be a family of unitary operators acting on $n+k$ qubits, \jonny{where $\jp=\{p_1,p_2,\dots\}$ is some set of parameters defining a unitary quantum circuit.} Also let $\ket{a}_{B'}$ be some fixed pure reference state of $k$ qubits. Using classical learning methods, we wish to find the unitary $U^\jp$ which maximizes the average fidelity, which we define to be the cost function,
\begin{equation}
C_1(\jp)=\sum_i p_i \cdot F(\ket{\psi_i},\rho_{i,\jp}^{out}), \label{eq:cost1}
\end{equation}
where,
\begin{equation}\label{eq:rhoout}
\rho_{i,\jp}^{out}=(U^\jp)^\dagger_{_{AB'}} \Tr_{_{B}} \bigg[ U^\jp_{_{AB}}\Big[\psi_{i_{AB}}\otimes a_{_{B'}}\Big](U^\jp_{_{AB}})^\dagger\bigg](U^\jp)_{_{AB'}}\; ,
\end{equation}
and we have abbreviated $\ket{\psi_i}\bra{\psi_i}_{AB}=\psi_{i_{AB}}$ and $\ket{a}\bra{a}_{B'}=a_{B'}$.
Equivalently, the goal is to find the best unitary $U^\jp$ which, on average, best preserves the input state of the circuit in \fig{generalcircuit} where instead of tracing over the $B$ system, we employ a swap gate and trace over the $B'$ system.  

To prove this, consider the fidelity of the input and output of the entire system of \fig{generalcircuit} for some fixed $\jp$ where we have denoted the swap operation by the unitary $V$,
\begin{eqnarray}
&{}& F(\ket{\psi_i}_{AB}\otimes \ket{a}_{B'},U^\dagger_{AB}V_{BB'}U_{AB}\ket{\psi_i}_{AB}\otimes \ket{a}_{B'})= \nonumber \\ 
&{}&F(U_{AB}\ket{\psi_i}_{AB}\otimes \ket{a}_{B'},V_{BB'}U_{AB}\ket{\psi_i}_{AB}\otimes \ket{a}_{B'})= \nonumber \\ 
&{}& F(\ket{\psi_i'}_{AB}\otimes \ket{a}_{B'},V_{BB'}\ket{\psi_i'}_{AB}\otimes \ket{a}_{B'})= \nonumber \\
&{}& F(\ket{\psi_i'}_{AB}\otimes \ket{a}_{B'},\ket{\psi_i'}_{AB'}\otimes \ket{a}_{B}),
\end{eqnarray}
where we have denoted $U\ket{\psi_i}=\ket{\psi_i'}$.  The terms in the cost function are found by tracing out over the $B'$ system,
\begin{eqnarray}
&{}& F(\Tr_{B'}[\psi'_{i_{AB}}\otimes a_{_B'}],\Tr_{B'} [\psi'_{i_{AB'}}\otimes a_{_B}])= \nonumber \\
&{}& F(\psi'_{i_{AB}},\rho'_{_A}\otimes a_{_B}),
\end{eqnarray}
where $\rho'_A=\Tr_{B'}\ket{\psi_i'}\bra{\psi_i'}_{AB'}]$.  
However, consider instead tracing over the $AB$ system and looking at the ``trash system'' of $B'$,
\begin{eqnarray}
&{}& F(\Tr_{AB}\big[\psi'_{i_{AB}}\otimes a_{_B'}\big],\Tr_{AB}\big[\psi'_{i_{AB'}}\otimes a_{_B}\big])= \nonumber \\
&{}& F(\ket{a}_{B'},\rho'_{B'}),
\end{eqnarray}
where $\rho'_{B'}=\Tr_A[\ket{\psi_i'}\bra{\psi_i'}_{AB'}]$.  We henceforth refer to $\rho'_{B'}$ as the ``trash state'' of the circuit.  It is straightforward to see in the above circuit that perfect fidelity (i.e. $C_1=1$) can be achieved by a unitary $U$ if and only if, for all $i$:
\begin{equation}
U|\psi_i\rangle_{AB} = |\psi^{c}_i\rangle_{A} \otimes |a\rangle_{B}\;. \label{eq:decoupling}
\end{equation}
where $\ket{\psi^{c}_i}_{A}$ is some compressed version of $\ket{\psi_i}$.  This follows because, if the $B$ and $B'$ systems are identical when the swap occurs, the entire circuit reduces to the identity map.  However, this occurs precisely when the trash state is equal to the reference state, i.e., $F(\ket{a}_{B'},\rho'_{B'})=1$.  This implies that it is possible to accomplish the learning task of finding the ideal $U^\jp$ by training \textit{only on the trash state}.  Furthermore, because \eq{decoupling} is completely independent of $U^\dagger$, this suggests that the circuit of \fig{generalcircuit} can be reduced further.  We then consider an alternative definition of the cost function in terms of the trash state fidelity,
\begin{equation}
C_2(\jp)=\sum_i p_i \cdot F(\Tr_A\Big[U^\jp\ket{\psi_i}\bra{\psi_i}_{AB}(U^\jp)^\dagger\Big],\ket{a}_B), \label{eq:cost2}
\end{equation}
Note, however, that the cost functions of \eq{cost1} and \eq{cost2} are not in general the same (in fact, $C_1\leq C_2$). \jonny{However, in practice, one must consider resource limitations; it is not hard to see that preparing copies of a fixed reference state would be easier than requiring identical copies of the input state to use in a SWAP test on the entire output state.  For some applications of a quantum autoencoder, it may also be the case that one has limited access to or limited knowledge of the input state.}

\begin{figure*}
\begin{tabular}{cccc}
\large{(a)}& \includegraphics[height=3.5cm]{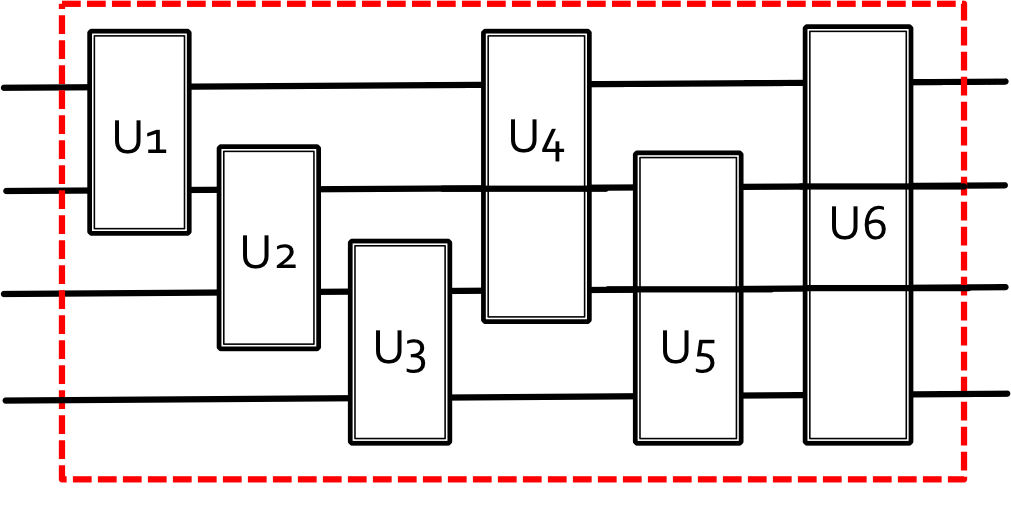} &
\large{(b)}& \includegraphics[height=3.5cm]{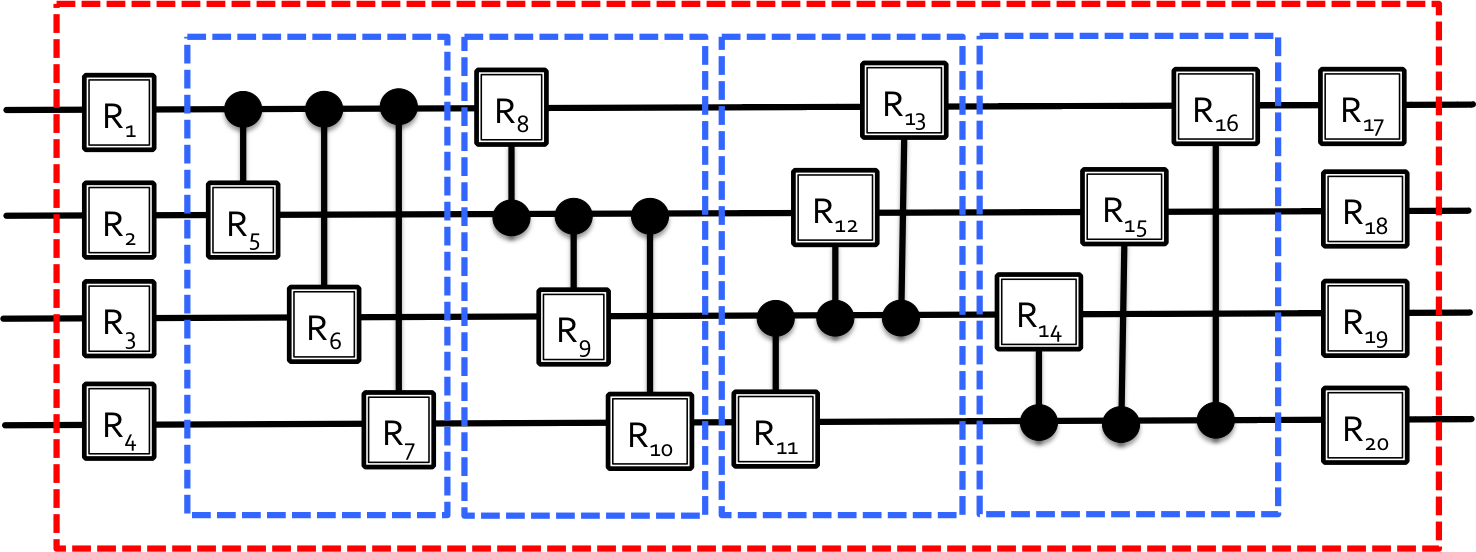}
\end{tabular}
\caption{Two programmable circuits employed as autoencoder models: a) Circuit A: a network of all the possible two-qubit gates (denoted by $\text{U}_i$) among the qubits. b) Circuit B: a network comprising all the possible controlled general single-qubit rotations (denoted by $\text{R}_i$) in a qubit set, plus single qubit rotations at the beginning and at the end of the unit-cell. All the circuits are depicted in the case of a four-qubit input. The unit-cell is delimited by the red dotted line.} \label{fig:heuristics}
\end{figure*}

It is interesting to note that, if we only care about circuits where $C_2\approx 1$, we can re-imagine the problem to being one of finding a particular disentangling.  \jonny{It has been shown that, employing a circuit of exponential depth, one is always able to perform a disentangling operation \cite{disentangle}, but to perform this operation in constant or polynomial depth is hard, and so classical heuristics are often used to find quantum circuits that are as close to optimal as possible.} Also, information-theoretic bounds have been explored in this context before, both in the context of one-shot compression and one-shot decoupling \cite{oneshotcompress,oneshotdecoupling}.  However, because the heuristics involved in choosing efficient-to-implement families of unitaries are largely ad-hoc, it is difficult to say if these bounds are meaningful in the context of a quantum autoencoder.

\section{Implementation of the quantum autoencoder model}\label{sec:implementation}

To implement the quantum autoencoder model on a quantum computer we must define the form of the unitary, $U^\jp$ (\eq{rhoout}) and decompose it into a quantum circuit suitable for optimization. For the implementation to be efficient, the number of parameters and the number of gates in the circuit should scale polynomially with the number of input qubits. This requirement immediately eliminates the possibility of using a $(n+k)$-qubit general unitary as $U^\jp$ due to the exponential scaling in the number of parameters needed to generate them.  

One alternative for the generation of $U^\jp$ is to employ a programmable quantum circuit \cite{sousa.arXiV.2006,daskin.JCP.137.234112.2012}. This type of circuit construction consists of a fixed networks of gates, \jrf{where a polynomial number of parameters associated to the gates i.e. rotation angles, constitute $\vec{p}$}. The pattern defining the network of gates is regarded as a unit-cell. This unit-cell can ideally be repeated to increase the flexibility of the model. For the numerical assessment presented in this work, we employed two simple programmable circuits illustrated in \fig{heuristics}. 

Circuit A has a unit-cell comprising a network of general two-qubit gates where we have considered all the possible pairings between qubits, as illustrated in \fig{heuristics}a for the four-qubit case. Accordingly, this model requires $15n(n-1)/2$ training parameters per unit-cell. A network of arbitrary two qubit gates can be easily implemented using state of the art superconducting qubit technologies \cite{Barends.N.508.500.2014} and the standard decomposition of a two-qubit gate into three CNOT gates and single-qubit rotations \cite{kraus.PRA.63.062309.2001}. Arbitrary two qubit-gates have been also implemented using ion traps \cite{hanneke.NP.6.13.2010} and quantum dots \cite{veldhorst.N.526.410.2015}.

Circuit B has a unit-cell comprising all the possible controlled one-qubit rotations among a set of qubits, complemented with a set of single qubit rotations at the beginning and at the end of the unit-cell, as shown in \fig{heuristics}b for the four-qubit case. We start considering the rotations controlled by the first qubit, followed by the rotations controlled by the second qubit and so on. Accordingly, our second model comprises $3n(n-1)+6n$ training parameters per unit-cell and can be implemented in state of the art quantum hardware using the standard decomposition of controlled unitaries into two CNOT gates and single-qubit rotations \cite{nielsen2010quantum}. This model is also general and can be modified by adding constraints to the parameters. For instance, one could consider the initial and final layers of rotations to be all the same.

\begin{figure}
\includegraphics[width=8.5cm]{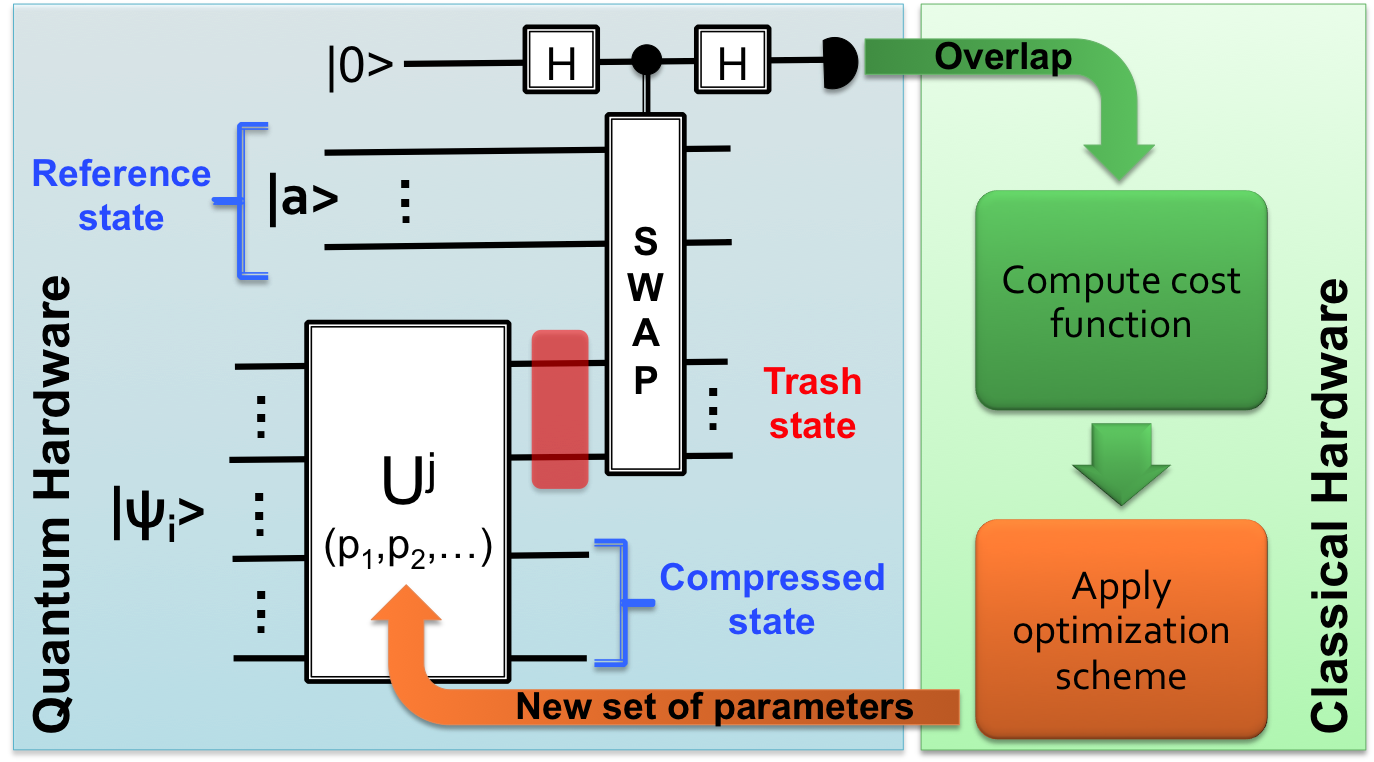}
\caption{Schematic representation of the hybrid scheme for training a quantum autoencoder. After preparation of the input state, $|\psi_i\rangle$, the state is compressed by the application of the parameterized unitary, $\text{U}^\jp$. The overlap between the reference state and the trash state produced by the compression is measured via a SWAP test. The results for all the states in the training set are collected to compute the cost function that is minimized using a classical optimization algorithm. The process is repeated until achieving  convergence on the cost function and/or the values of the parameters, $\vec{p} = (\text{p}_1,\text{p}_2,\dots)$.} \label{fig:training}
\end{figure}

Once the circuit model has been chosen, we must train the network by maximizing the autoencoder cost function \eq{cost2}, in close analogy to classical autoencoders. Our training procedure adopts a quantum-classical hybrid scheme in which the state preparation and measurement are performed on the quantum computer while the optimization is realized via an optimization algorithm running on a classical computer. Such hybrid schemes have been proposed in the context of quantum machine learning \cite{gammelmark.NJP.11.033017.2009,bang.NJP.16.073017.2014} and variational algorithms for quantum simulation \cite{Peruzzo.NC.5.4213.2014,Mcclean.NJP.18.023023.2016,Wecker.PRA.92.042303.2015,Li.variational.arXiV.2016}. In the later case, several experimental demonstrations have been successfully carried out \cite{Peruzzo.NC.5.4213.2014,omalley.PRX.6.031007.2016,santagati.arXiV.2016}.

As described in \sec{model}, the cost function of the quantum autoencoder is defined as the weighted average of fidelities between the trash state produced by the compression,  and the reference state. These fidelities can be measured via a SWAP test \cite{buhrman.PRL.87.167902.2001} between the reference state and the trash state. Accordingly, our quantum register must comprise the input state, $|\psi_i\rangle$, and the reference state. In a single iteration of our training algorithm, we perform the following steps for each of the states in the training set:
\begin{enumerate}
\item Prepare the input state, $|\psi_i\rangle$, and the reference state. We assume these preparations to be efficient.
\item Evolve under the encoding unitary, $U^\jp$, where $\vec{p}$ is the set of parameters at a given optimization step.
\item Measure the fidelity between the trash state and the reference state via a SWAP test.
\end{enumerate}
With the estimates of all the fidelities, the cost function (\eq{cost2}) is computed and fed into a classical optimization routine that returns a new set of parameters for our compression circuit. These steps are repeated until the optimization algorithm converges. Given that the value of the cost function is upper bounded by 1, we performed the optimization by minimizing the value of the function $\log_{10}\left(1-C_2\right)$. This procedure is widely used in machine learning applications and helps prevent numerical instabilities \cite{bishop2006}. A graphical summary of the hybrid scheme for training a quantum autoencoder is shown in \fig{training}.

\begin{figure}
\includegraphics[width=8.0cm]{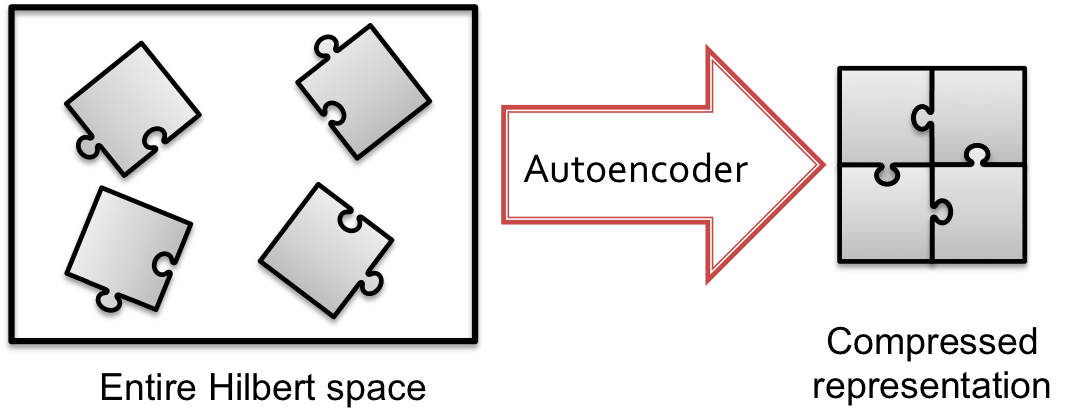}
\caption{Graphical representation of Hilbert space compression. Given that the states of interest have support on only a subset $\mathcal{S}$ of the Hilbert space (gray pieces), the quantum autoencoder finds an encoding that uses a space of size $|\mathcal{S}|$.}\label{fig:hilbertSpaceCompression}
\end{figure}

\begin{figure}
\includegraphics[width=8.0cm]{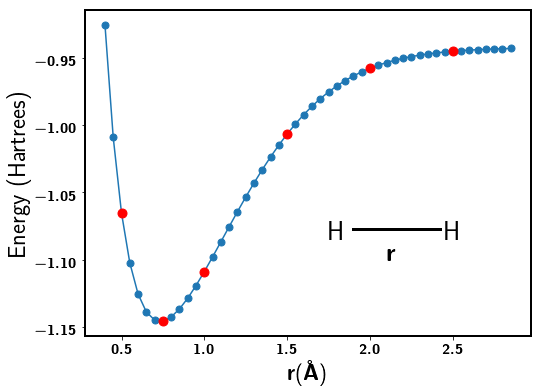}
\caption{Potential energy surface for the hydrogen molecule using a STO-6G basis set. The ground states at the red dots where used as training set for the quantum autoencoder. The ground states at the blue dots were used for testing.}\label{fig:h2pes}
\end{figure}

\section{Application to quantum simulation}

Consider a set of states, $\{|\psi_i\rangle\}$, with support on a subset  of a Hilbert space $\mathcal{S} \subset \mathcal{H}$. Using a quantum autoencoder, we could find an encoding scheme that employs only $\log_2|\mathcal{S}|$ qubits to represent the states instead of $\log_2|\mathcal{H}|$, with a trash state of size $\log_2|\mathcal{H}-\mathcal{S}|$. This idea is graphically depicted in \fig{hilbertSpaceCompression}. This situation is usually encountered for eigenstates of many-body systems due to special symmetries. 

Fermionic wavefunctions, for instance, are eigenfunctions of the particle number operator, same as the fermionic state vectors. Consequently, an eigenstate of a system with $\eta$ particles is spanned exclusively by the subspace of fermionic state vectors with the same number of particles \cite{Helgaker2013}, that has size ${N \choose \eta}$ with $N$ the number of fermionic modes. This result has direct implications for the design of quantum algorithms for simulation, suggesting that the number of qubits required to store fermionic wavefunctions could be reduced up to $\log {N \choose \eta} $ if an appropriate mapping can be found. The same situation is encountered for the spin projection operator, thus reducing the size of the subspace spanning a specific fermionic wavefunction even further. 

Generally, the number of particles of the system is part of the input when finding eigenstates of many-body systems. In quantum chemistry simulations, the spin projection of the target state is also known. Many classical algorithms for simulating quantum systems take advantage of these constraints to reduce their computational cost \cite{Helgaker2013}. However, the standard transformations employed to map fermionic systems to qubits, namely the Bravyi-Kitaev (BK) and the Jordan-Wigner (JW) mappings \cite{Seeley.JCP.137.224109.2012,Tranter.115.1431.IJQC.2015}, do not exploit these symmetries and thus employ more qubits than formally needed. 

In this scenario, a quantum autoencoder could be trained to compress fermionic wavefunctions obtained using a quantum simulation algorithm that has been run using the standard transformations. The compression schemes obtained through this procedure could be employed to reduce the use of quantum memory, if the wavefunction needs to be stored. It also could save quantum resources for the simulation of systems with similar symmetries. To illustrate this idea, we simulated a quantum autoencoder applied to molecular wavefunctions.

\jrf{ Within the Born-Oppenheimer approximation, the non-relativistic molecular Hamiltonian can be written as
\begin{align}\label{eq:molHamiltonian}
H=h_{nuc}+\sum_{pq}h_{pq} a^{\dagger}_p a_q + \frac{1}{2} \sum_{pqrs} h_{pqrs} a^{\dagger}_p a^{\dagger}_q a_r a_s
\end{align}
where $h_{nuc}$ corresponds to the classical electrostatic repulsion between nuclei, and the constants $h_{pq}$ and $h_{pqrs}$ correspond to the one- and two-electron integrals (see \app{molecularHam}). The operators $a^{\dagger}_p$ and $a_p$ creates and annihilates an electron in the spin-orbital $p$. After applying either the JW or the BK transformation, the molecular Hamiltonian can be expressed as $H=\sum^M_i c_i H_i$, with $M$ scaling as $O(N^4)$. In this case, the operators $H_i$ correspond to tensor products of Pauli matrices and the real coefficient $c_i$ are linear combinations of the one- and two-electron integrals. For a fixed set of nuclei and a given number of electrons, the molecular integrals as well as the coefficients $c_i$ are functions of the internal coordinates of the molecule, $\vec{R}$.}

For instance, consider the Hamiltonian of molecular hydrogen in the STO-6G minimal basis set \cite{Helgaker2013}. Using the JW transformation, the corresponding Hamiltonian acting on four qubits adopts the generic form \cite{Seeley.JCP.137.224109.2012}:
\begin{equation}\label{eq:hamH2}
\begin{aligned}
H &= c_0 I + c_1 (Z_0 + Z_1) + c_2 ( Z_2 + Z_3) + c_3 Z_0Z_1 + \\
& c_4( Z_0Z_2 + Z_1Z_3) + c_5( Z_1Z_2 + Z_0Z_3) + c_6 Z_2Z_3 \\
& + c_7( Y_0X_1X_2Y_3 - X_0X_1Y_2Y_3 - Y_0Y_1X_2X_3 + X_0Y_1Y_2X_3)
\end{aligned}
\end{equation}

In this case, the coefficients $c_i$ are a function of the internuclear distance, $r$. By solving the Schr\"odinger equation for the Hamiltonians at different values of $r$, we can obtain the ground state energy for molecular hydrogen and construct the potential energy surface (PES) shown in \fig{h2pes}. We expect that the ground state wavefunctions along the PES conserve the same number of particles and projection spin symmetries, turning this set of states into an excellent target for compression.

To illustrate the previous idea, we classically simulated a quantum autoencoder taking six ground states of the hydrogen molecule at different values of $r$, $\{|\Psi(r_i)\rangle \}_{i=1}^{6}$, as our training set. In this case, the weights of the states are chosen to be all equal. In real applications, we can imagine that the ground states are obtained using a quantum algorithm such as the quantum variational eigensolver \cite{Peruzzo.NC.5.4213.2014}. We trained the circuit model described in \fig{heuristics} to compress the training set of four-qubit states to two qubits and to one qubit, using $|0\rangle^{\otimes 2}$ and $|0\rangle^{\otimes 3}$ as reference states, respectively. Once the circuits were trained we tested them on 44 ground states corresponding to values of $r$ different from those of the training set. This selection of the training and testing sets is shown in \fig{h2pes}. 

\begin{table}[h]
\caption{Average fidelity ($\mathcal{F}$) error after one cycle of compression and decompression using the quantum autoencoder trained from ground states of the Hydrogen molecule. We also report the error in the energy of the decoded state. (Maximum and minimum errors shown within parenthesis). 6 states were used for training and 44 more were used for testing. These results were obtained with the L-BFGS-B optimization.}\label{tab:H2fidelities}
\begin{tabular}{ccccc}
\hline
\specialcell{Circuit} & \specialcell{Final size\\ (\# qubits)} & Set & \specialcell{$-\log_{10}(1-\mathcal{F})$ \\ MAE} & \specialcell{-log$_{10}$ Energy \\ MAE\\(Hartrees)} \\
\hline
Model	&	2	&	Training	&	6.96(6.82-7.17)	&	6.64(6.27-7.06)	\\
A		&	2	&	Testing	&	6.99(6.81-7.21)	&	6.76(6.18-7.10)	\\
		&	1	&	Training	&	6.92(6.80-7.07)	&	6.60(6.23-7.05)	\\
		&	1	&	Testing	&	6.96(6.77-7.08)	&	6.72(6.15-7.05)	\\
\hline
Model	&	2	&	Training	&	6.11(5.94-6.21)	&	6.00(5.78-6.21)	\\
B		&	2	&	Testing	&	6.07(5.91-6.21)	&	6.03(5.70-6.21)	\\
		&	1	&	Training	&	3.95(3.53-5.24)	&	3.74(3.38-4.57)	\\
		&	1	&	Testing	&	3.81(3.50-5.38)	&	3.62(3.35-4.65)	\\
\hline
\multicolumn{5}{l}{$^*$ {\footnotesize MAE: Mean Absolute Error. Log chemical accuracy in Hartrees $\approx$-2.80 }}
\end{tabular}
\end{table}

\begin{figure}
\includegraphics[width=8.0cm]{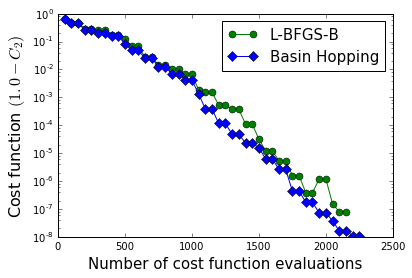}
\caption{A plot of the cost function versus the number of cost function evaluations during the training process. This example corresponds to a quantum autoencoder for compression of wavefunction of H$_2$ from 4 to 2 qubits using circuit A with a training set of six ground states. We compared the L-BFGS-B and the Basin-Hopping algorithms for optimization.}\label{fig:optimization}
\end{figure}

The classical simulation was performed using a Python script supplemented with the QuTiP library \cite{johansson.CPC.183.1760.2012,johansson.CPC.184.1234.2013}. To simulate general two-qubit gates we employed the decomposition described in Ref.\cite{hanneke.NP.6.13.2010}. Arbitrary single-qubit rotations were implemented by decomposing them into Pauli-Z and Pauli-Y rotations, $R =R_z(\theta_1)R_y(\theta_2)R_z(\theta_3)$, ignoring global phases \cite{nielsen2010quantum}. The optimization was performed using the SciPy implementation of the Basin-Hopping (BS) algorithm \cite{wales.JPCA.101.5111.1997}. We also employed the L-BFGS-B method \cite{byrd.SIAMJSC.16.1190.1995} with a numerical gradient (central finite difference formula with step size $h=10^{-8}$). In the optimization of both circuit models, the parameters were constrained to the range $[0,4\pi)$. The optimization of each circuit was initialized by setting the parameters to randomly selected values.

\tab{H2fidelities} shows the average error in the fidelities and the energies obtained after a cycle of compression and decompression through the optimal quantum autoencoder. We observe that both circuit models are able to achieve high fidelities for the encoding, producing decoded wavefunctions with energies that are close to the original values within chemical accuracy ($1\text{kcal/mol} \equiv 1.6\times10^{-3} \ \text{Hartrees} \ \equiv 43.4 \ \text{meV}$). This accuracy requirement assures that quantum chemistry predictions have enough quality to be used for the estimation of thermochemical properties such as reaction rates \cite{peterson.TCA.131.1.2012}.

\newcommand{\model}{latentA}
\newcommand{\modelb}{latentB}
\newcommand{\myH}{3.5cm}
\begin{figure*}
\begin{tabular}{ccccc}
\hline
&& \normalsize{$r=0.75\AA$} & \normalsize{$r=1.50\AA$} & \normalsize{$r=2.50\AA$} \\ 
\hline
\rotatebox{90}{\quad \quad \large{Input space}} & \rotatebox{90}{\quad \quad \quad \large{(4 qubits)}} & \includegraphics[height=\myH]{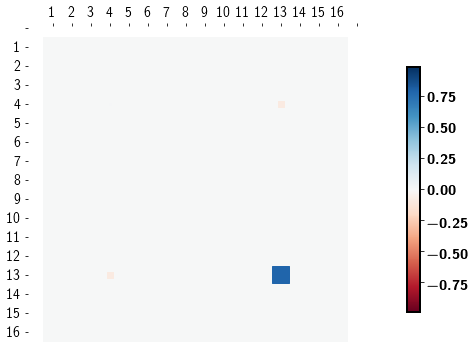} & \includegraphics[height=\myH]{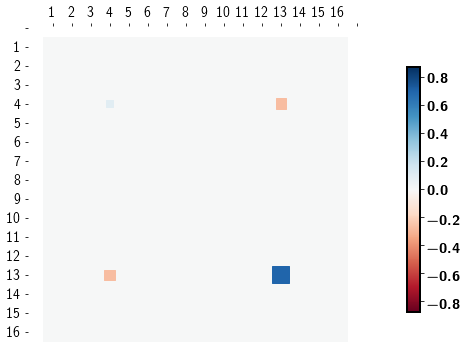} & \includegraphics[height=\myH]{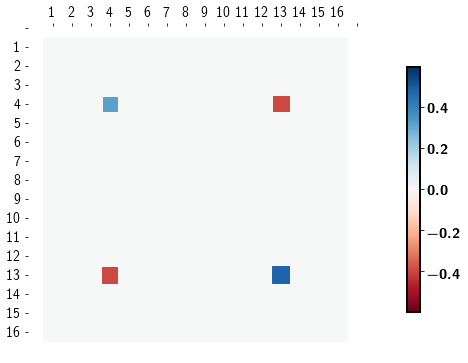} \\
\hline
\rotatebox{90}{\quad \quad \large{Latent space (A)}} & \rotatebox{90}{\quad \quad \quad \large{(2 qubits)}} & \includegraphics[height=\myH]{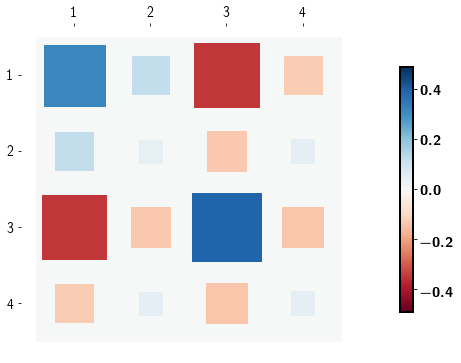} & \includegraphics[height=\myH]{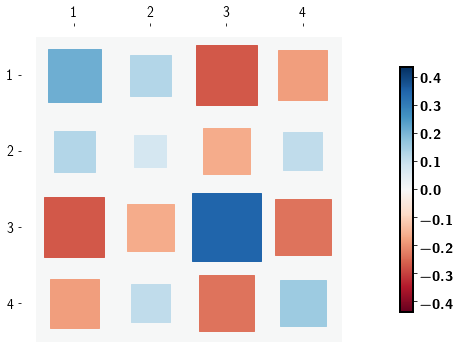} & \includegraphics[height=\myH]{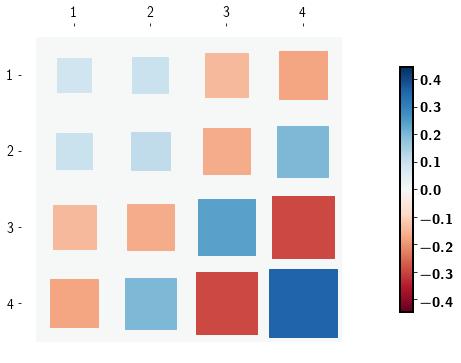} \\
\hline
\rotatebox{90}{\quad \quad \large{Latent space (B)}} & \rotatebox{90}{\quad \quad \quad \large{(2 qubits)}} & \includegraphics[height=\myH]{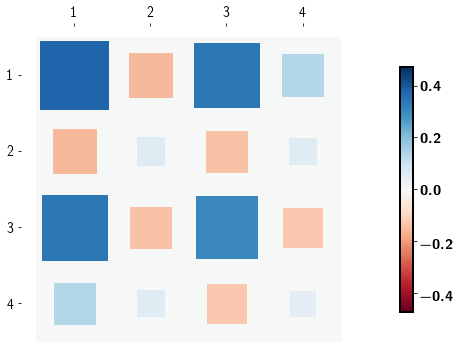} & \includegraphics[height=\myH]{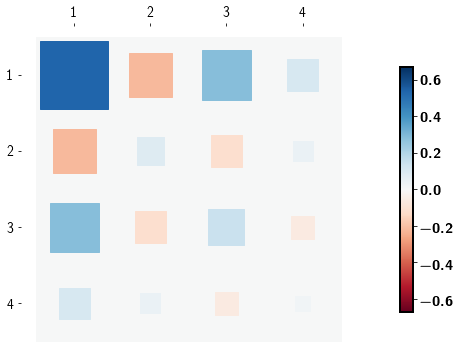} & \includegraphics[height=\myH]{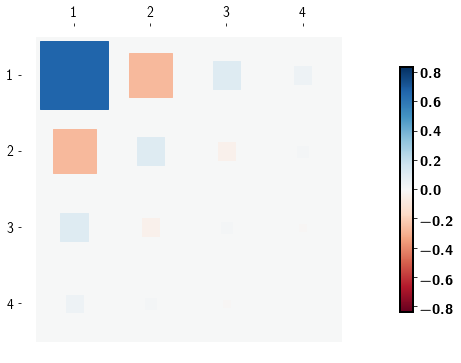} \\
\hline
\rotatebox{90}{\quad \quad \large{Latent space (A)}} & \rotatebox{90}{\quad \quad \quad \large{(1 qubit)}} & \includegraphics[height=\myH]{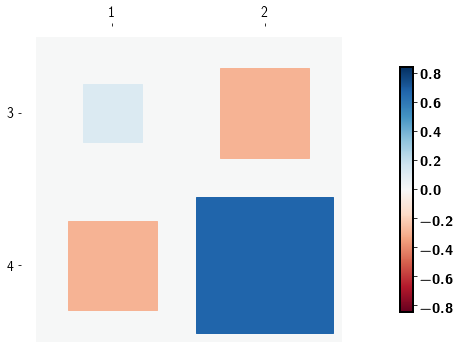} & \includegraphics[height=\myH]{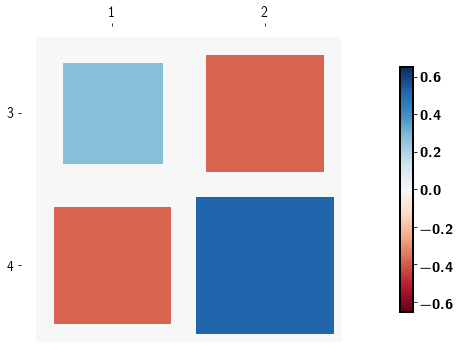} & \includegraphics[height=\myH]{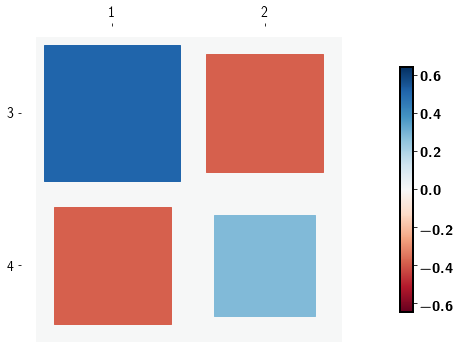} \\
\hline
\rotatebox{90}{\quad \quad \large{Latent space (B)}} & \rotatebox{90}{\quad \quad \quad \large{(1 qubit)}} & \includegraphics[height=\myH]{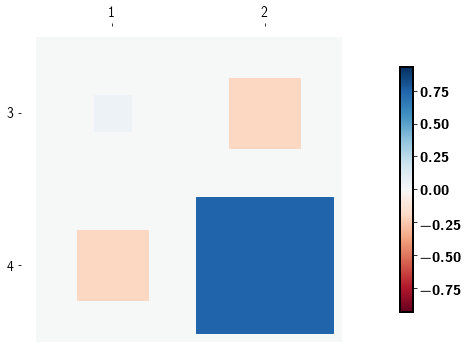} & \includegraphics[height=\myH]{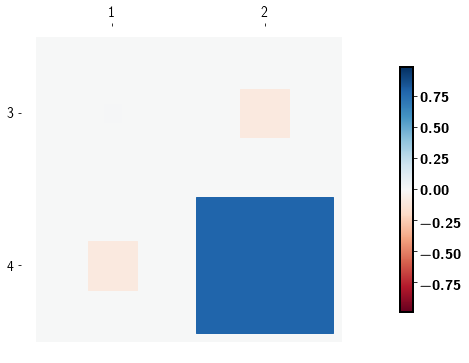} & \includegraphics[height=\myH]{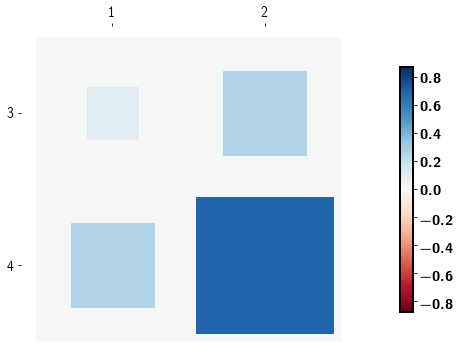} \\
\hline
\end{tabular}
\caption{Visualization of the input space and the latent (compressed) spaces for three different testing instances of the H$_2$ compression, corresponding to three different bond distances, $r$. The input and latent spaces are characterized as the density matrices of the input and compressed states. Letters (A) and (B) denote the type of circuit employed to construct the encoding unitary. The size of the register (in qubits) appears within parenthesis. Integer labels starting at 1 denote the computational basis states in ascending order from $|00\cdots0\rangle$ to $|11\cdots1\rangle$.} \label{fig:latentSpaceA}
\end{figure*}

\fig{optimization} illustrates the optimization process of a quantum autoencoder. We compared two different optimization algorithms, L-BFGS-B and Basin-Hopping. The parameters were initialized at random and the same guess was employed for both optimizations. As observed in \fig{optimization}, both algorithms required a similar number of cost function evaluations to achieve similar precision and exhibit a monotonic reduction of the difference between the cost function and its ideal value with the number of function evaluations. The implementation of the quantum autoencoder in state of the art architectures can benefit from the use of algorithms that do not require gradient evaluations and have a larger tolerance to the presence of noise in the hardware, such as Basin-Hopping. 

To gain insight into the compression process, we plotted the density matrices of the compressed states and compared them with the density matrix of the original state in  \fig{latentSpaceA}, for three different values of $r$. The sparsity of the original input density matrix is due to the symmetry of the Hamiltonian for molecular hydrogen, whose eigenvectors have support on only 2 computational basis states, allowing for a compression up to a single qubit. Although the quantum autoencoder achieves high fidelity with both types of circuit, the structure of the density matrices indicates that the forms of the compressed space and therefore the forms of the encoding unitaries differ between the two circuit models. As the values of $r$ change, the relation between the features of the input space, here represented by the elements of the density matrix, and the features of the compressed space become apparent. 

\begin{figure}
\includegraphics[width=3.0cm]{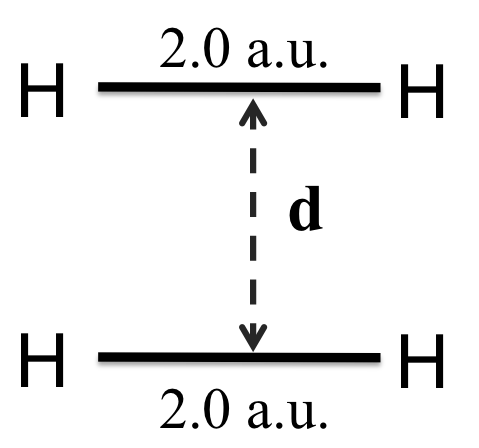}
\caption{H$_4$ molecule in a parallel configuration, with the hydrogen atoms forming a rectangle. We obtained the ground states of this system at different values of $\rm{d}$, with the bond distance in the two hydrogen molecules fixed to 2 atomic units (a.u.).} \label{fig:H4}
\end{figure}

\begin{table}[h]
\caption{\jrf{Final error in the cost function (\eq{cost2}) obtained after training a quantum autoencoder for compression of the ground states of two-sites and four-sites Hubbard models and the H$_4$ molecule along a parallel path. Six ground states were used for training each system. These results were obtained with the L-BFGS-B optimization.}}\label{tab:otherSystems}
\begin{tabular}{cccc}
\hline
\specialcell{Circuit}	& 	System 	&	\specialcell{ Compression rate$^*$ \\ $n_o \to n_l$}			&	 \specialcell{$-\log_{10}(1-C_2)$}	\\
\hline									
Model	&	Hubbard	&	4	$\to$	3	&	7.52	\\
A	&	2x1 sites	&	4	$\to$	2	&	1.15	\\
	&		&	4	$\to$	1	&	1.13	\\
\cline{2-4}									
	&	Hubbard	&	8	$\to$	7	&	2.28	\\
	&	2x2 sites	&	8	$\to$	6	&	1.42	\\
\cline{2-4}									
	&	H$_4$	&	8	$\to$	7	&	1.53	\\
	&		&	8	$\to$	6	&	1.6	\\
\hline									
Model	&	Hubbard	&	4	$\to$	3	&	6.82	\\
B	&	2x1 sites	&	4	$\to$	2	&	3.92	\\
	&		&	4	$\to$	1	&	4.02	\\
\cline{2-4}									
	&	Hubbard	&	8	$\to$	7	&	2.29	\\
	&	2x2 sites	&	8	$\to$	6	&	2.31	\\
\cline{2-4}									
	&	H$_4$	&	8	$\to$	7	&	4.33	\\
	&		&	8	$\to$	6	&	1.15	\\
\hline									
\multicolumn{4}{l}{$^*$ {\footnotesize $n_o$ and $n_l$ stand for the number of qubits in the original register}} \\ 
\multicolumn{4}{l}{{\footnotesize and the number of qubits in the latent space, respectively.}}
\end{tabular}
\end{table}

\jrf{In addition to the example of molecular hydrogen, we tested the autoencoder compression scheme with ground states of the Hubbard model and the H$_4$ molecule. We considered half-filled Hubbard models with 2-sites and 4-sites arranged in a two-leg ladder (2$\times$1 and 2$\times$2 lattices, respectively). The Hamiltonian for these systems is given by
\begin{equation}
H=-t\sum_{<i,j>}\sum_{\sigma} a^{\dagger}_{i,\theta} a_{j,\sigma} + U \sum_{i} a^{\dagger}_{i,\uparrow} a_{i,\uparrow} a^{\dagger}_{i,\downarrow} a_{i,\downarrow}
\end{equation}
where $a^{\dagger}_{i,\theta}$ and $a_{i,\sigma}$ create and annihilate an electron at site $i$ with spin $\sigma$, respectively. The summation in the first term runs over all the interacting sites, denoted as $\textless i,j \textgreater$. We used periodic boundary conditions along the horizontal direction and open boundary conditions in the vertical direction. As in the case of molecular Hamiltonians, Hubbard Hamiltonians can be mapped to qubits using either the JW or the BK transformation, requiring two qubits per site. 

We trained the two circuits of \fig{heuristics} to compress the ground states of the Hubbard Hamiltonians obtained by setting $t$ to six different values equally spaced between 0.9 and 1.1, with $U=2.0$. The optimization process was repeated three times starting at randomly selected values. The same procedure was applied to the ground states of the H$_4$ system at six different values of the bond distance $d$ (0.6, 1.4, 2.2, 3.0, 3.8 and 4.6 atomic units) for the geometry described in \fig{H4}.

\tab{otherSystems} shows the lowest errors obtained for the compression of the Hubbard models and the H$_4$ system. Errors are quantified as the difference between the final value of the cost function (\eq{cost2}) and the ideal value of 1. Recall that the cost function corresponds to the average fidelity over the training set. We observe that the ground states of the two-sites Hubbard model can be compressed from 4 to 3 qubits using both circuit types. However, only circuit B is able to compress these these states from 4 to 2 qubits and 4 to 1 qubits with an error below $10^{-3}$. The same circuit-type achieves an error smaller than $10^{-4}$ when compressing the ground states of the H$_4$ system from 8 to 7 qubits. In contrast, circuit A is unable to obtain errors below $10^{-3}$ for H$_4$. In the case of the 4-sites Hubbard model, none of the circuit models proposed here was able to obtain errors below $10^{-3}$. 

The differences between the performances of the two circuit models described above exemplifies how the ansatz employed for the autoencoder unitary impacts the degree of compression achievable with the autoencoder model. Compression of a particular set of states could be achieved more easily with a dedicated ansatz designed for that purpose. One form of unitary that can serve as a template to design such dedicated ansatzes is given by the expression
\begin{equation}\label{eq:Uham}
U(\vec{\alpha})=e^{-i\sum_{i} \alpha_i H_i}
\end{equation}
where the real numbers $\alpha_i$ are the parameters for optimization and the terms $H_i$ are local interactions consisting of tensor products of Pauli matrices. The experimental implementation of \eq{Uham} would benefit from the techniques developed for quantum simulation algorithms \cite{Georgescu.RMP.86.153.2014}.

Finally, we point out that the maximum rate of lossless compression achievable with a quantum autoencoder is predetermined by the size of the subspace spanning the training set. Consequently, a given set of states might only admit a small or null compression rate. For instance, consider a fermionic system with 8 fermionic modes and 4 particles, such as a half-filled 4-sites Hubbard model or the H$_4$ molecule in a minimal basis set studied here. Based solely on the constrain in the number of particles, these 8-qubits systems could be compressed to $\log_2 {8 \choose 4} \approx 7$ qubits. Compression beyond this point could be achieved if an extra symmetry constraint is present. In general, we expect fermionic systems where the number of fermionic modes is considerably larger than the number of particles to be good candidates for compression.
}

\section{Discussion}

We have introduced a general model for a quantum autoencoder -- a quantum circuit augmented with learning via classical optimization -- and have shown that it is capable of learning a unitary circuit which can facilitate compression of quantum data, particularly in the context of quantum simulations.  We imagine that the model can have other applications, such as compression protocols for quantum communication, error-correcting circuits, or perhaps to solve particular problems directly. \jrf{A natural application for quantum autoencoders is state preparation. Once a quantum autoencoder has been trained to compress a specific set of states, the decompression unitary ($U^{\dagger}$) can be used to generate states similar to those originally used for training. This is achieved by preparing a state of the form $\ket{\Psi_I} \otimes \ket{a}$ and evolving it under $U^{\dagger}$, where $\ket{\Psi_I}$ has the size of the latent space and $\ket{a}$ is the reference state used for training. 

Autoencoders as state preparation tools have direct application in the context of quantum variational algorithms \cite{Peruzzo.NC.5.4213.2014,Mcclean.NJP.18.023023.2016,Wecker.PRA.92.042303.2015,Li.variational.arXiV.2016}. These algorithms approximate the energy or the time evolution of an eigenstate by performing measurements on a quantum state prepared according to a parameterized ansatz. A quantum autoencoder could be trained with states of size $n_o$ qubits, obtained from a given ansatz, and later be used as a state preparation tool as described above. Because the autoencoder parameters are fixed after training, the only active parameters in the variational algorithm would be those associated to the preparation of a state in the latent space ($n_l$). Since $n_l<n_o$, the state preparation with autoencoders would require fewer parameters than the original ansatz.
}

\jonny{In our specification of the autoencoder, we define the input states to be an ensemble of pure states, and the evolution of those states to be unitary.  The most generalized picture of the autoencoder, however, would allow for inputs to be ensembles of mixed states and the set $\{U^\jp\}$ to be a set of quantum channels. In the case of mixed state inputs, we remark that this formulation can in principle be captured by our model already.  More specifically, one can consider the case where a set of ancillas (denoted $A'$) representing a purification of the mixed state is input into the autoencoder along with the original input. Ulhmann's theorem \cite{Wilde2012} guarantees that there exists a purification whose fidelity is identical to that of the mixed state generated from tracing out the purification; namely, it is a maximum over a unitary $V$ acting on the purification alone (although finding this unitary can be a difficult computational problem itself).  Consider then the encoding $U^\jp_{AB}\otimes V_{A'}$, where the original latent space is expanded to contain all of $A'$ (i.e. none of the ancilla qubits are traced out).   This purified system will recover the behavior of the mixed system.  The autoencoder structure as defined here cannot completely capture the structure for general quantum channels, though we expect other computational tasks may be solved by considering specific channel instances.}


Are there any obvious limitations to the quantum autoencoder model?  One consideration is that the von-Neumann entropy \cite{Wilde2012} of the density operator representing the ensemble $\{p_i,\ket{\psi_i}_{AB}\}$ limits the number of qubits to which it can be noiselessly compressed.  However, finding the entropy of this density operator is not trivial -- in fact, given a circuit that constructs a density operator $\rho$, it is known that, in general, even estimating the entropy of $\rho$ is QSZK-complete \cite{QEA}.  This then opens the possibility for quantum autoencoders to efficiently give \textit{some} estimate of the entropy of a density operator.  \jonny{In a similar vein, the unitary of the autoencoder could be defined to include the action of a quantum channel, and the autoencoder used to give both an encoding for and some lower bound for the capacity of a quantum communication channel (although the trash state may no longer be useful for training the autoencoder in some of these cases).}

It is natural to consider whether the quantum autoencoder structure we have defined is actually a generalization of its classical counterpart, as in the construction of \cite{FFNN2016}.  It may certainly be possible that some particular family of unitaries, together with certain choices for $n$ and $k$, can be constructed so that a mapping exists.  However, it is unclear that such a correspondence would even be desirable.  Rather, we believe the value of autoencoders in general lies in the relatively simple structure of forcing a network to preserve information in a smaller subspace, as we have defined here.

Another topic of interest for any quantum computing model is the computational complexity exhibited by the device.  For our construction, it is clear that any complexity result would be dependent upon the family of unitaries that is chosen for the learning to be optimized over.  As the training itself is based on classical optimization algorithms (with no clear `optimal' learning method), this further obfuscates general statements regarding the complexity of the model.


\section*{Acknowledgements}

The authors would like to thank Yudong Cao, Peter Johnson, Ian Kivlichan, Nicolas Sawaya, Libor Veis, and Dominic Berry for very insightful discussions and suggestions. JR and AAG acknowledge the Air Force Office of Scientific Research for support under Award: FA9550-12-1-0046. AA-G and JO acknowledge support from the Vannevar Bush Faculty Fellowship program sponsored by the Basic Research Office of the Assistant Secretary of Defense for Research and Engineering and funded by the Office of Naval Research through grant N00014-16-1-2008.
AA-G acknowledges the Army Research Office under Award: W911NF-15-1-0256. The authors thank the Harvard Odyssey cluster facility where the numerical simulations presented in this work were carried out.


\appendix

\section{Molecular integrals}\label{app:molecularHam}

Using atomic units, where the electron mass $m_e$, the electron charge $e$, Bohr radius $a_0$, Coulomb's constant and $\hbar$ are unity, we can write the nuclear repulsion, $h_{nuc}$, and one-electron and two-electron integrals, $h_{pq}$ and $h_{pqrs}$, as
\begin{equation}
h_{pq} = \int d\sigma \varphi_p^*(\sigma) \left(-\frac{\nabla_{\vec{r}}^2}{2} - \sum_{i} \frac{Z_i}{|\vec{R}_i - \vec{r}|} \right)\varphi_q(\sigma) \label{eq:single_int}
\end{equation}
\begin{equation}
h_{pqrs} = \int d\sigma_1\ d\sigma_2\ \frac{ \varphi_p^*(\sigma_1) \varphi_q^*(\sigma_2)  \varphi_s(\sigma_1) \varphi_r(\sigma_2) }{|\vec{r}_1 - \vec{r}_2|} \label{eq:double_int}
\end{equation}
\begin{equation}
h_{nuc}=\frac{1}{2}\sum_{i \neq j} \frac{Z_i Z_j}{|\vec{R}_i-\vec{R}_j|}
\end{equation}
Where $Z_i$ represents the nuclear charge, $\vec{r}$ and $\vec{R}$ denote electronic and nuclear spatial coordinates, respectively, and $\sigma$ is now a spatial and spin coordinate with $\sigma_i=(\vec{r}_i;s_i)$. Summations run over all nuclei. $\varphi(\sigma)$ represent the spin-orbitals (one-electron functions), that are generally obtained from a self-consistent field (SCF) Hartree-Fock (HF) calculation.


\bibliographystyle{apsrev}
\bibliography{biblio}

\end{document}